\documentclass[twocolumn]{aastex631}
\usepackage{graphicx}
\usepackage{subfigure}
\usepackage{amsmath}
\usepackage{epsfig}

\newcommand{\imgdir}{./images_submit}
\newcommand{\refsec}[1]{Section~\ref{#1}}
\newcommand{\reffig}[1]{Figure~\ref{#1}}
\newcommand{\refeq}[1]{Equation~(\ref{#1})}

\newcommand{\TNG}{IllustrisTNG}

\shorttitle{Angular Momentum - Mass Relation}
\shortauthors{Du et al.}

\begin{document}
\title{The Origin of the Relation Between Stellar Angular Momentum and Stellar Mass in Nearby Disk-dominated galaxies} 

\correspondingauthor{Min Du}
\email{dumin@xmu.edu.cn}

\author{Min Du}
\affil{Department of Astronomy, Xiamen University, Xiamen, Fujian 361005, China}

\author{Luis C. Ho}
\affiliation{Kavli Institute for Astronomy and Astrophysics, Peking University, Beijing 100871, China}
\affiliation{Department of Astronomy, School of Physics, Peking University, Beijing 100871, China}

\author{Hao-Ran Yu}
\affil{Department of Astronomy, Xiamen University, Xiamen, Fujian 361005, China}

\author{Victor P. Debattista}
\affiliation{Jeremiah Horrocks Institute, University of Central Lancashire, Preston PR1 2HE, UK}

\begin{abstract}
The \TNG\ simulations reproduce the observed scaling 
relation between stellar specific angular momentum (sAM) $j_{\rm s}$
and mass $M_{\rm s}$ of central galaxies. We show that the local 
$j_{\rm s}$-$M_{\rm s}$ relation 
${\rm log}\ j_{\rm s} = 0.55 \ {\rm log}\ M_{\rm s} + 2.77$ 
develops at $z\lesssim 1$ in disk-dominated 
galaxies. We provide a simple model that 
describes well such a connection between halos and galaxies. 
The index 0.55 of the $j_{\rm s}$-$M_{\rm s}$ relation comes 
from the product of the indices of the $j_{\rm tot}\propto M_{\rm tot}^{0.81}$, 
$M_{\rm tot}\propto M_{\rm s}^{0.67}$, and 
$j_{\rm s}\propto j_{\rm tot}$ relations, where 
$j_{\rm tot}$ and $M_{\rm tot}$ are overall sAM and mass of a halo. 
A non-negligible deviation from the tidal torque theory, which predicts 
$j_{\rm tot}\propto M_{\rm tot}^{2/3}$, should be included. This model 
further suggests that the stellar-to-halo mass ratio 
of disk galaxies increases monotonically 
following a nearly power-law function that is consistent
with the latest dynamical measurements. 
Biased collapse, in which galaxies form from the inner and 
lower sAM portion of their parent halos, has a minor effect 
at low redshifts. The retention factor of angular momentum 
reaches $\sim 1$ in disk galaxies with strong rotations, 
and it correlates inversely with the mass fraction of 
the spheroidal component, which partially explains 
the morphological dependence of the $j_{\rm s}$-$M_{\rm s}$ 
relation. 

\end{abstract}
\keywords{Scaling relations (2031); Galaxy kinematics (602); Galaxy evolution (594); Spiral galaxies (1560); Galaxy dark matter halos (1880)}

\section{Introduction}

The relation between the properties of galaxies and 
their parent dark matter halos, as well as the physical 
processes that regulate such properties, is a long-standing 
puzzle. In a cosmological framework, the angular momentum 
of a dark matter halo is initially acquired 
through tidal torques from neighbouring perturbations. 
The classical tidal torque theory \citep{Hoyle1949, Peebles1969} 
predicts that the specific angular momentum $j_{\rm h}$ 
(sAM hereafter) of halos follows $j_{\rm h}\propto M_{\rm h}^{2/3}$ where 
$M_{\rm h}$ is the halo mass \citep[e.g.][]{Peebles1969, White1984}.
If angular momentum is 
conserved throughout the formation of galaxies by 
accreting gas that decoupled from their host dark 
matter halos, a similar relation should also apply 
to galaxies. Observations show that stellar masses 
$M_{\rm s}$ and sAM $j_{\rm s}$ of disk galaxies 
are correlated as a power law with index $0.52-0.64$ 
\cite[e.g.,][]{Fall1983, Romanowsky&Fall2012, Fall&Romanowsky2013, Posti2018b, DiTeodoro2021, Hardwick2022, Pina2021}. 
This empirical trend is often called the ``Fall relation''. The model
$j_{\rm s} \propto f_j f_{\rm m}^{-2/3} M_{\rm s}^{2/3} \lambda$ 
\citep{Romanowsky&Fall2012} has been widely used to explain the 
$j_{\rm s}$-$M_{\rm s}$ relation. It requires that the retention 
factor of angular momentum $f_j\equiv j_{\rm s}/j_{\rm h}$, 
the stellar-to-halo mass ratio $f_{\rm m} \equiv M_{\rm s}/M_{\rm h}$, 
and the spin parameter $\lambda$ are independent of stellar mass. 
Unless these conditions are met, their dependence on stellar mass 
must conspire to cancel out to generate a correlation of the form 
$j_{\rm s} \propto M_{\rm s}^{2/3}$.

One of the key ingredients of the galaxy-halo connection is 
$f_{\rm m}$ or the the stellar-to-halo mass relation 
\citep[SHMR; reviewed by][]{Wechsler&Tinker2018}. Within the 
general framework of abundance matching, the stellar-to-halo 
mass ratio peaks in halos around $10^{12} M_\odot$, assuming there is a 
little or no dependence on galaxy morphology. This result 
directly leads to a non-linear $j_{\rm s}$-$M_{\rm s}$ 
relation in logarithmic space, which is inconsistent 
with observations, as 
discussed by \citet{Posti2018a}. Yet, several works 
suggest that the exact shape of the SHMR is not 
independent of galaxy morphology 
\citep[e.g.,][]{Mandelbaum2006, Dutton2010, Rodriguez-Gomez2015, Posti&Fall2021}. 
Recently, \citet{Posti2019a, Posti2019b}, using a 
sample of isolated disk galaxies with presumably more 
accurate halo masses measured dynamically, found that 
the SHMR follows a nearly linear relation in logarithmic 
space \citep[see][for some extremely massive cases]{DiTeodoro2021, DiTeodoro2022}. 
\citet{ZhangZhiwen2022} reached a similar result for 
more than 20000 star-forming galaxies whose dynamical masses 
were measured via galaxy-galaxy lensing and satellite kinematics.

Another key ingredient is the retention factor of angular momentum $f_j$.
Disky structures grow at $z \lesssim 2$ by accreting cold gas 
from the vast reservoir of their circumgalactic medium 
\citep[e.g.,][]{Tacchella2019, DeFelippis2020, Renzini2020, Du2021}. 
During this phase, the angular momentum of the gas should be 
conserved (up to a certain factor) to form disk galaxies 
with angular momenta tightly correlated with that of 
their parent dark matter halos. But no agreement is fully 
reached in studies that examined the link of the angular 
momentum amplitude between halos and galaxies. \citet{Zavala2016} 
and \citet{Lagos2017} found a remarkable connection between the 
sAM evolution of the dark and baryonic components of 
galaxies in the EAGLE simulations. A similar 
correlation is suggested in \citet{Teklu2015} using the 
Magneticum Pathfinder simulation. \citet{Grand2017} and 
\citet{Rodriguez-Gomez2022} showed that the 
disk sizes and scale lengths are closely related to the 
angular momentum of halos in the Auriga and IllustrisTNG-100 
simulations. However, \citet{Jiang2019} found little to no 
correlation using the NIHAO zoom-in simulation.
A similar conclusion was drawn in \citet{Scannapieco2009} using 
eight Milky Way analogs. \citet{Danovich2015}
argued that cold gas inflows cannot conserve angular momentum when 
they move into the inner regions of halos.  

In this paper, we use \TNG\ 
\citep{Marinacci2018, Nelson2018a, Nelson2019a, Naiman2018, Pillepich2018b, Pillepich2019, Springel2018} to revisit the longstanding 
open question of how the $j$-$M$ relation develops in disk galaxies . 
We aim to address: (1) whether or not there is a connection 
between the angular momentum of dark halos and that of the 
galaxies they host; (2) how the $j_{\rm s}$-$M_{\rm s}$ 
evolves in disk galaxies; and (3) how the
$j_{\rm s}$-$M_{\rm s}$ relation can be explained using a simple
theoretical model. 

\section{TNG50 Simulation and data reduction}

\TNG\ is a suite of cosmological simulations 
that was run with gravo-magnetohydrodynamics and incorporates a 
comprehensive galaxy model \citep[][]{Weinberger2017, Pillepich2018a}. 
The TNG50-1 run of the \TNG\ has the highest resolution. It includes 
$2 \times 2160^3$ initial resolution elements in a $\sim 50$ 
comoving Mpc box, corresponding to a baryon mass resolution of 
$8.5 \times 10^4 M_\odot$ with a gravitational softening length for 
stars of about $0.3$ kpc at $z = 0$. Dark matter is resolved with 
particles of mass $4.5 \times 10^5 M_\odot$. Meanwhile, the minimum 
gas softening length reaches 74 comoving pc. This resolution is 
able to reproduce the kinematic properties of galaxies with 
stellar mass $\gtrsim 10^9 M_\odot$ \citep{Pillepich2019}. The 
galaxies are identified and characterized with the 
friends-of-friends \citep{Davis1985} and 
{\tt SUBFIND} \citep{Springel2001} algorithms. Resolution elements 
(gas, stars, dark matter, and black holes) belonging to an 
individual galaxy are gravitationally bound to its host subhalo. 

In this work, we mainly focus on how the $j_{\rm s}$-$M_{\rm s}$ 
relation develops in central galaxies dominated by disks. In such 
cases, neither mergers nor environmental effects have played an 
important role. We use the galaxies over the stellar mass range 
$10^{9}-10^{11.5} M_\odot$ from the TNG50-1 run. Disk-dominated 
galaxies are identified by $\kappa_{\rm rot} \geq 0.5$, where 
$\kappa_{\rm rot} = K_{\rm rot}/K$ \citep{Sales2012} denotes the relative 
importance of cylindrical rotational energy $K_{\rm rot}$ over 
the total kinetic energy $K$ measured for a given snapshot. 
\citet{Du2021} showed that 
$\kappa_{\rm rot} \geq 0.5$ selects galaxies whose mass fractions 
of kinematically derived spheroidal structures are $\lesssim 0.5$. 
The other galaxies are classified as spheroid-dominated galaxies, 
which correspond to elliptical galaxies or slow rotators in 
observations. We further divide disk-dominated galaxies into two 
subgroups with $\kappa_{\rm rot} \geq 0.7$ and 
$0.5 \leq \kappa_{\rm rot} <0.7$, which correspond to the 
cases with strong rotation and with relatively moderate 
rotation, respectively, for a given snapshot. 
The former ones are likely to have more disky morphology. 

All quantities in this paper are calculated using all particles 
belonging to galaxies/subhalos that include all 
gravitationally bound particles identified with the 
{\tt SUBFIND} algorithm \citep{Springel2001}. We  
use only central galaxies that are primary 
subhalos of their parent halos. 
The specific angular momentum vector is thus 
${\boldsymbol j} = \sum_{i} {\boldsymbol J_{i}}/\sum_{i} m_{i}$, 
where ${\boldsymbol J_{i}}$
and $m_{i}$ are the angular momentum and mass of particle $i$, 
respectively. Galaxies are centered at the position with the 
minimum gravitational potential energy. No limitation on the 
radial extent is made to obtain the overall properties. The 
radial variation is ignored to simplify our discussion. 

\section{The generation of the \lowercase{$j_{\rm s}$}-$M_{\rm \lowercase{s}}$ relation of disk galaxies at \lowercase{$z=0$}}

\begin{figure*}
\begin{center}
\includegraphics[width=0.98\textwidth]{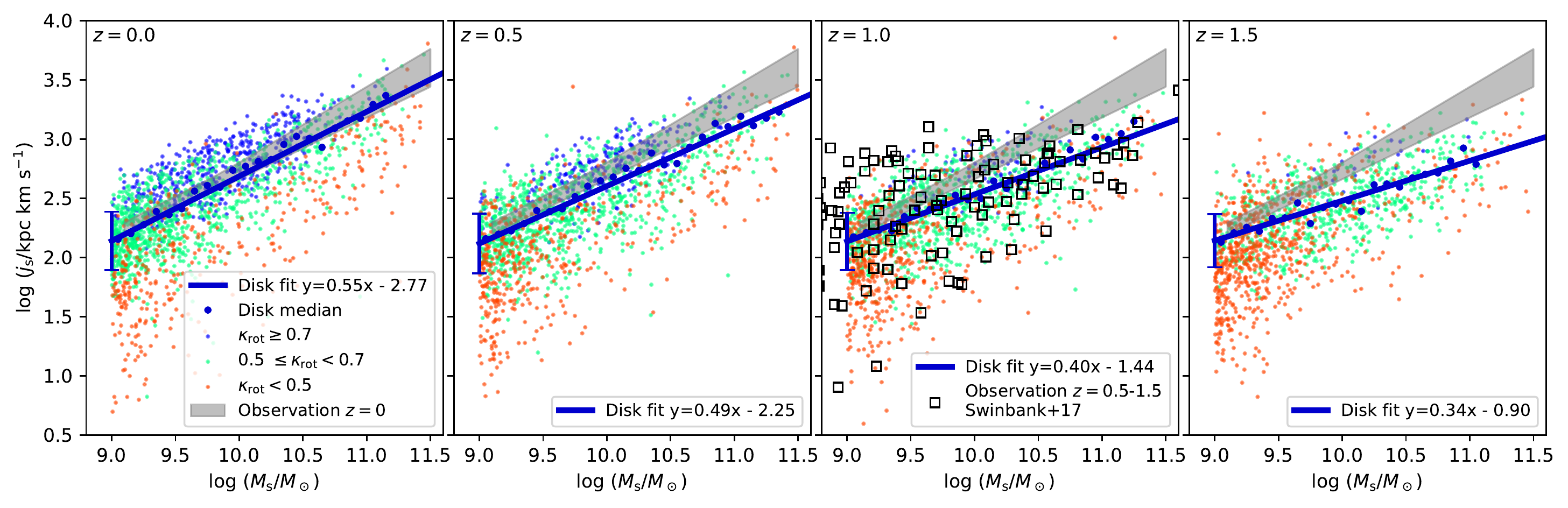}
\caption{The evolution of the $j_{\rm s}$-$M_{\rm s}$ relation in TNG50 from $z=1.5$ to $z=0$. The red, green, and blue symbols are central galaxies that correspond to spheroid-dominated galaxies with $\kappa_{\rm rot} < 0.5$, disk-dominated galaxies with $0.5 \leq \kappa_{\rm rot} < 0.7$, and disk-dominated galaxies with $\kappa_{\rm rot} \geq 0.7$, respectively. The redshift is given at the top-left corner of each panel. The blue lines with error bars are the linear fitting results of disk-dominated galaxies. The error bars represent the standard deviation from the linear fitting, which is 0.3 dex at $z=0$. The large blue dots show the trend of the median values. The shaded region shows the variance of the $j_{\rm s}$-$M_{\rm s}$ relation suggested by observations at $z=0$, where we combine the fitting results given by \citet{Romanowsky&Fall2012}, \citet{Fall&Romanowsky2013}, \citet{Posti2018b}, \cite{DiTeodoro2021}, and \citet{Pina2021}. In the third panel, the black squares show the $j_{\rm s}$-$M_{\rm s}$ relation measured for disk galaxies at $z=0.5-1.5$ \citep{Swinbank2017}.}
\label{fig:Msjs_evo}
\end{center}
\end{figure*}

In the left-most panel of \reffig{fig:Msjs_evo}, we show 
the $j_{\rm s}$-$M_{\rm s}$ relation of galaxies at $z=0$ from 
TNG50, in comparison with those measured in observations. 
The shaded region encloses the fitting results of 
disk galaxies measured in the local Universe 
\citep{Romanowsky&Fall2012, Fall&Romanowsky2013, Posti2018b, Hardwick2022, Pina2021}. These studies concluded that $j_{\rm s}$-$M_{\rm s}$ 
follows a well-defined linear scaling relation in logarithmic 
space with slope $0.52-0.64$ and a root-mean-square 
scatter of $\sim 0.2$ dex. It is clear that TNG50 reproduces well
the $j_{\rm s}$-$M_{\rm s}$ relation observed in disk-dominated 
central galaxies (small blue and cyan dots). A linear fit 
(blue line) of all disk-dominated galaxies of TNG50 gives 
\begin{equation}\label{eqjsMs}
	\begin{aligned}
    {\rm log}\ j_{\rm s} = (0.55\pm 0.01)\ {\rm log}\ M_{\rm s} - (2.77\pm 0.11),
    \end{aligned}
\end{equation}
with a scatter ($0.3$ dex) similar to that observed. 
In this study, we adopt the linear regression and 
the least-squares method in fitting. The median trend 
(large blue dots) that matches the linear fitting well. 
We focus on the general trend and physical origin 
of the $j_{\rm s}$-$M_{\rm s}$ relation.  

Figure \ref{fig:Msjs_evo} further shows that the $j_{\rm s}$-$M_{\rm s}$ 
relation in the local Universe develops at $z\lesssim 1$, 
which coincides well with the epoch of the formation and growth 
of disk galaxies. Its slope becomes shallower at higher redshifts, 
thus deviating from the $j_{\rm s}$-$M_{\rm s}$ relation at $z=0$ 
(shaded regions); for example, 
log $j_{\rm s} = 0.34\ {\rm log}\ M_{\rm s} - 0.90$ at $z=1.5$. 
In the third panel of \reffig{fig:Msjs_evo}, we can see 
that the disk galaxies at $z=0.5-1.5$ measured by 
\citet{Swinbank2017} follow a consistent distribution with 
the TNG50 disk galaxies. It is worth mentioning again that 
here the galaxies for each redshift are 
kinematically classified by $\kappa_{\rm rot}$.
At high redshifts, the spheroid-dominated galaxies follow a 
similar $j_{\rm s}$-$M_{\rm s}$ relation as the disk-dominated cases, 
but with a larger scatter. As less disk-dominated galaxies form at 
higher redshifts, this result suggests that the growth of disky 
structures at late times ($z < 1$) is the key to establishing the 
locally observed $j_{\rm s}$-$M_{\rm s}$ relation in disk galaxies.  

The decrease of $j_{\rm s}$ toward high redshifts is most likely 
due to the effect of biased collapse \citep[e.g.,][]{vandenBosch1998}, which 
predicts that gas with less angular momentum collapses earlier. 
The formation of galaxies at high redshifts thus is largely 
dominated by the assembly of spheroidal components whose angular 
momentum correlates weakly with that of their parent halos. 
The effect of biased collapse is gradually weakened toward low 
redshifts due to the assembly of disks by accretion of 
gas with high angular momentum. In this study, we focus 
mainly on the generation of the $j_{\rm s}$-$M_{\rm s}$ 
relation in disk-dominated galaxies at low redshifts. The effect 
of biased collapse is examined later in the paper.

\section{A physical model of the \lowercase{$j_{\rm s}$}-$M_{\rm \lowercase{s}}$ relation}

The existence of the $j_{\rm s}$-$M_{\rm s}$ relation 
suggests that, despite the complexity of galaxy formation 
in a cosmological context, a fundamental regularity 
still exists. This relation can be directly obtained 
by three simple equations:
\begin{equation}\label{eqjM}
	\begin{aligned}
	{\rm log}\ j_{\rm tot} = \alpha\ {\rm log}\ M_{\rm tot} + a
	\end{aligned}
\end{equation}
\begin{equation}\label{eqMM}
	\begin{aligned}
	{\rm log}\ M_{\rm tot} = \beta\ {\rm log}\ M_{\rm s} + f_m^{'}
	\end{aligned}
\end{equation}
\begin{equation}\label{eqjj}
	\begin{aligned}
	{\rm log}\ j_{\rm s} = \gamma\ {\rm log}\ j_{\rm tot} + f_j^{'}.
	\end{aligned}
\end{equation}
Here $M_{\rm tot}$ and $j_{\rm tot}$ are the total mass 
and angular momentum of a halo system, including 
baryonic and dark matter. This model yields 
\begin{equation}\label{eqmain}
\begin{aligned}
{\rm log}\ j_{\rm s} = \alpha \beta \gamma\ {\rm log}\ M_{\rm s} + a\gamma + \alpha \gamma f_{\rm m}^{'} + f_j^{'}.
\end{aligned}
\end{equation}
Equation \ref{eqjM} is a general form of the theoretical 
prediction of the halo $j$-$M$ relation. Tidal 
torque theory suggests $\alpha=2/3$, but if we allow for 
potential correction to the theory, 
$\alpha$ may deviate from $2/3$. As suggested by 
\citet[][PFM19 hereafter]{Posti2019a}, we 
assume that the stellar-to-halo mass ratio follows a 
single power-law relation (i.e., equation \ref{eqMM}) 
for disk galaxies. We apply \refeq{eqjj} to describe the 
retention of angular momentum; the retention factor 
$f_j^{'}={\rm log}\ f_j$ if $\gamma=1$. 

In this section, we apply this simple model to the 
TNG50 data to show that they provide a good 
interpretation of the $j_{\rm s}$-$M_{\rm s}$ relation. 
\refsec{sec:fj} shows the angular momentum correlation 
between halos and stars and then examines 
whether the effect of biased collapse is important. 
In \refsec{sec:SHMR}, we show the SHMR and the $j$-$M$ 
relation of halos, which play important roles in establishing
the $j_{\rm s}$-$M_{\rm s}$ relation. It is worth emphasizing that 
our results are based on semi-quantitative analysis that is not 
sensitive to minor deviations from the scaling relations. 
All linear fitting results are roughly consistent with 
the trends of median values over the mass range considered. We 
further discuss how 
our results challenge the SHMR obtained by the abundance 
matching method and the halo $j$-$M$ relation predicted 
by tidal torque theory. 

\subsection{Angular momentum conservation during disk assembly}
\label{sec:fj}

\begin{figure*}
\begin{center}
\includegraphics[width=0.98\textwidth]{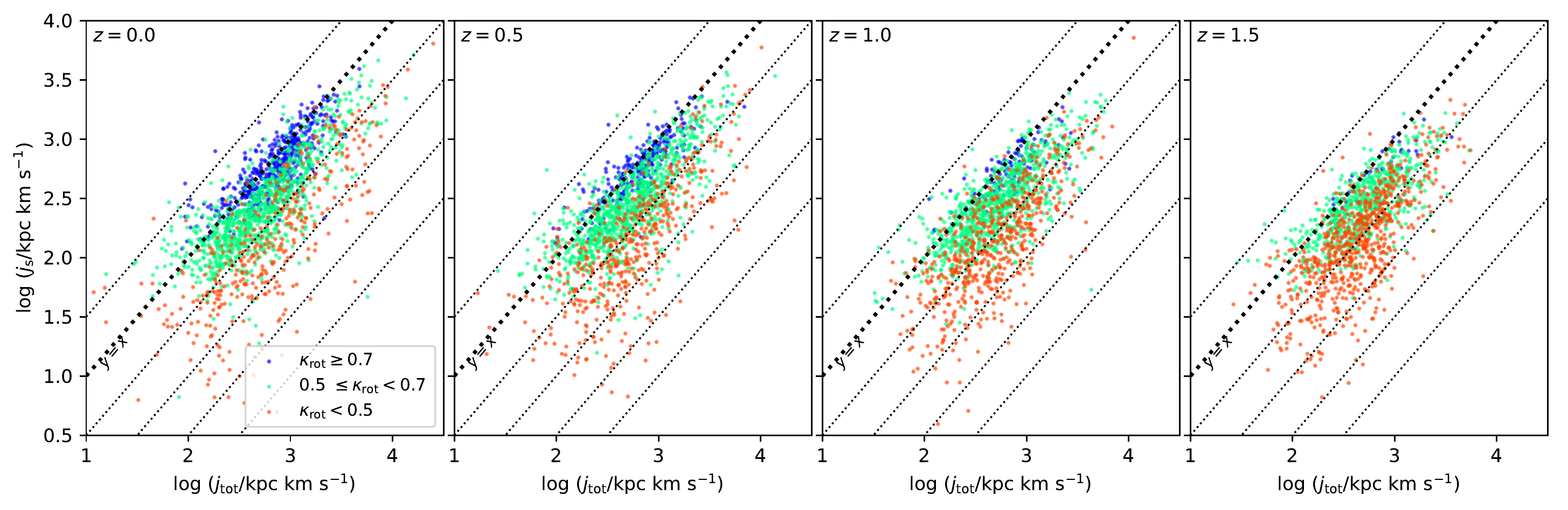}
\caption{Evolution of the $j_{\rm s}$-$j_{\rm tot}$ relation of central galaxies in TNG50. The red, green, and blue symbols are central galaxies that correspond to spheroid-dominated galaxies with $\kappa_{\rm rot} < 0.5$, disk-dominated galaxies with $0.5 \leq \kappa_{\rm rot} < 0.7$, and disk-dominated galaxies with $\kappa_{\rm rot} \geq 0.7$, respectively. The dotted lines highlight the log $j_{\rm s}={\rm log}\ j_{\rm tot} + f_j^{'}$ scaling relation in an interval of $\Delta f_j^{'} = 0.5$.}
\label{fig:fj}
\end{center}
\end{figure*}

\begin{figure}
\begin{center}
\includegraphics[width=0.5\textwidth]{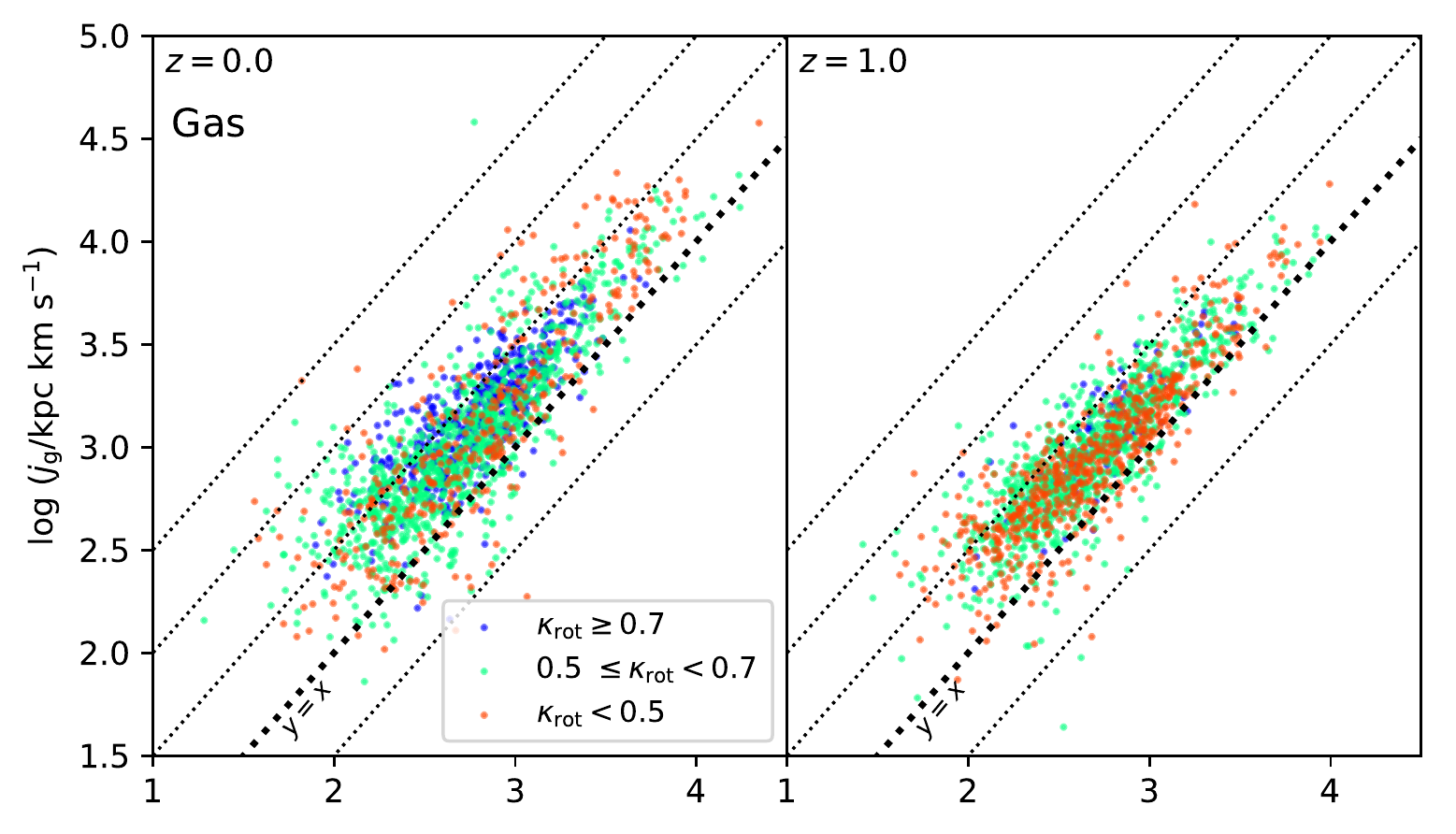}
\includegraphics[width=0.5\textwidth]{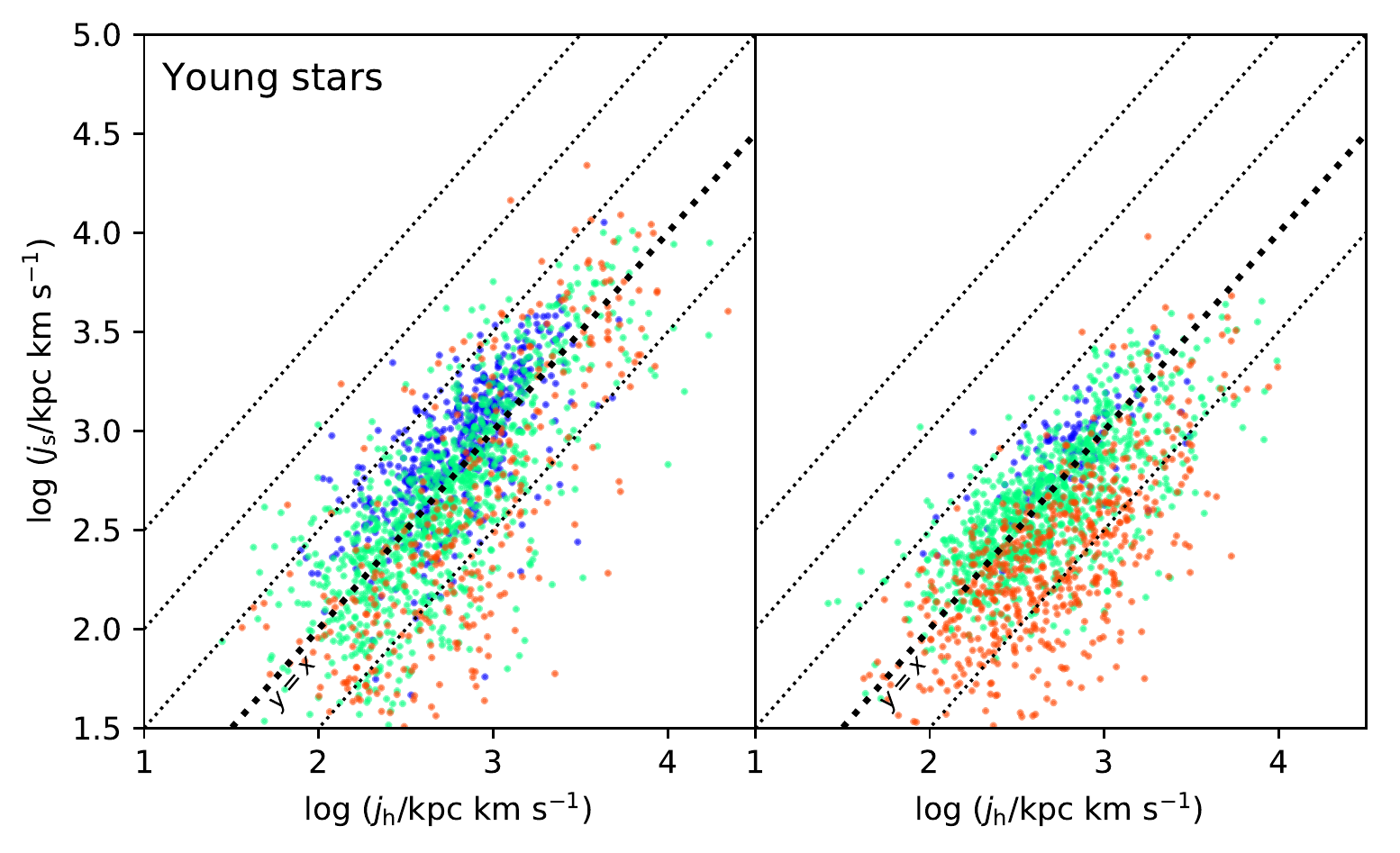}
\caption{The evolution of the $j$-$j_{\rm h}$ relation for gas (top) and young stars (bottom) in TNG50 at $z=0$ (left) and $z=1.0$ (right). The red, green, and blue symbols are central galaxies that correspond to spheroid-dominated galaxies with $\kappa_{\rm rot} < 0.5$, disk-dominated galaxies with $0.5 \leq \kappa_{\rm rot} < 0.7$, and disk-dominated galaxies with $\kappa_{\rm rot} \geq 0.7$, respectively. The dotted lines highlight the scaling relation in an interval of $0.5$ dex. We exclude the cases with star formation rates lower than $0.1\ M_\odot\ {\rm yr}^{-1}$ in the last 1 Gyr (i.e., quenched galaxies), which only contribute a small fraction ($\sim 1/4$ at $z=0$) of even the spheroid-dominated galaxies.}
\label{fig:jysjdm}
\end{center}
\end{figure}

\begin{figure}
\begin{center}
\includegraphics[width=0.45\textwidth]{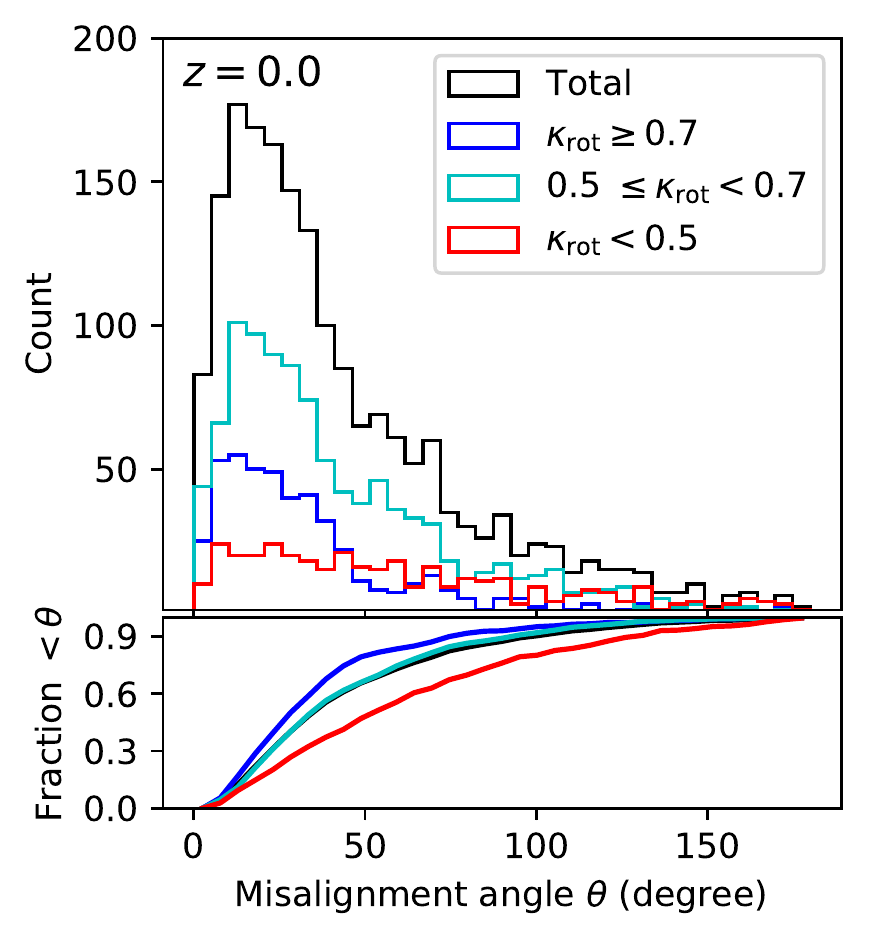}
\caption{The number distribution of misalignment angle $\theta$ (top) and its accumulative fraction (bottom) at $z=0$. The misalignment angle $\theta$ measures the angle between vectors ${\boldsymbol j_{\rm s}}$ and ${\boldsymbol j_{\rm h}}$. The three groups of galaxies are shown in blue, cyan, and red. The black curve corresponds to the distribution of all central galaxies.}
\label{fig:misangle}
\end{center}
\end{figure}

Previous works have suggested many phenomena that may induce angular 
momentum losses or gains, including dynamical friction, 
hydrodynamical viscosity, galactic winds 
\citep[e.g.,][]{Governato2007, Brook2011}, and galactic fountains 
\citep[e.g.,][]{Brook2012, DeFelippis2017}. These processes, 
in conjunction with gas cooling and subsequent star formation, 
drive circulation of gas in the circumgalactic medium.  

The tight correlation between $j_{\rm s}$ and $j_{\rm tot}$ in 
central disk-dominated galaxies at $z=0$ (\reffig{fig:fj}) 
verifies that the overall angular momentum is retained in 
a nearly constant ratio during star formation and gas circulation. 
This result supports the long-standing assumption 
from theory \citep[e.g.,][]{Fall&Efstathiou1980, Mo1998}  
and recent cosmological simulations \citep{Teklu2015, Zavala2016, Lagos2017} that 
angular momentum is approximately conserved during galaxy formation.
The galaxies with $\kappa_{\rm rot} \geq 0.7$
match the $y=x$ line (thick dotted line), which suggests 
that angular momentum is conserved in galaxies 
with strong rotation, giving $j_{\rm s} \sim j_{\rm tot}$, namely 
$\gamma \sim 1$ and $f_j^{'}\sim 0$. The disk-dominated galaxies with 
$0.5 \leq \kappa_{\rm rot} < 0.7$ are slightly offset parallel 
to $y=x$. Equation \ref{eqjj} can thus be written as 
${\rm log}\ j_{\rm s} \simeq {\rm log}\ j_{\rm tot} + f_j^{'}$,
where the offset $f_j^{'}$ decreases with $\kappa_{\rm rot}$ 
following a nearly parallel sequence. An accurate calculation of 
median retention factors gives $f_j^{'} = -0.07_{-0.17}^{+0.15}$ and
$-0.21_{-0.24}^{+0.20}$ for disk-dominated galaxies 
with $\kappa_{\rm rot} \geq 0.7$ and $\kappa_{\rm rot} \geq 0.5$ , 
respectively, which is consistent with the observational 
estimation for disk galaxies 
\citep{Fall&Romanowsky2013, Fall&Romanowsky2018, Posti2019b, DiTeodoro2021}.
For comparison, the spheroid-dominated galaxies 
(median $f_j^{'} = -0.63_{-0.36}^{+0.31}$) follow a  
weak correlation with a rather large scatter. They thus 
cannot be described by a linear relation.

It is worth emphasizing that the evolution of $j_{\rm s}$ is 
a cumulative effect that quantifies the overall conservation 
of angular momentum over the past evolution. In \reffig{fig:jysjdm}, 
we further show the relation between sAM of the dark matter halo 
($j_{\rm h}$) and that of gas ($j_{\rm g}$) and young stars. 
It is clear that $j_{\rm g}$ (upper panels) 
correlates linearly with $j_{\rm h}$, 
but offsets toward higher sAM by about $0$-$0.5$ dex, in 
qualitatively agreement with observations \citep{ManceraPina2021b}. 
In the lower panels of \reffig{fig:jysjdm}, we can see that 
$j_{\rm s}$ for the young stars and $j_{\rm h}$ of disk-dominated 
galaxies (especially the cases with $\kappa_{\rm rot} \geq 0.7$) 
follow roughly a similar linear scaling relation as 
the $j_{\rm g}$-$j_{\rm h}$ relation at $z=0$, 
albeit with a larger scatter and relatively lower sAM. Here 
$j_{\rm s}$ of young stars is approximated using stars that 
form within 1 Gyr in each galaxy for a given snapshot. 
This result suggests that the sAM of gas and the assembly 
of disks are largely determined by the sAM of their parent halos. 

The fact that gas and young stars have higher sAM than dark 
matter can be partially explained by the 
biased collapse scenario \citep[e.g.,][]{vandenBosch1998}. 
In this scenario, gas with lower angular momentum collapses 
earlier, whereupon the remaining gas, and consequently the young 
stars that form from it at lower redshifts, would be left with 
somewhat higher angular momentum. The conservation of sAM 
evidenced by \reffig{fig:fj} suggests, however, that the 
overall effect of biased collapse has been quite modest in 
disk-dominated galaxies after a sufficiently long period 
of gas accumulation. 
A dramatic loss of angular momentum only occurs in 
spheroid-dominated galaxies, probably due to  
dry major mergers that can destroy the global 
rotation of their initial disks. 

We further verify that the angular momentum vectors of 
the dark matter halo and stars are roughly 
aligned. Defining the misalignment angle $\theta$ as the angle 
between vectors ${\boldsymbol j_{\rm s}}$ and ${\boldsymbol j_{\rm h}}$, 
the lower panel of Figure \ref{fig:misangle} shows  
that $\sim 60\%$ of disk-dominated galaxies have 
$\theta < 30^{\circ}$ at $z=0$. This may induce a 
scatter on ${\boldsymbol j_{\rm h}}$ by a factor of 
$< 1-{\rm cos}\ 30^{\circ}=0.13$, which 
is negligible. This result is consistent with previous studies 
\citep[e.g.,][]{Bailin2005, Bett2010, Teklu2015, ShaoShi2016}. 
\citet{MotlochYu2021} further 
find a correlation between galaxy spin direction and  
halo spin reconstructed from cosmic initial conditions
\citep{YuHaoRan2020, WuQiaoYa2021}. We thus ignore the 
effect of orientation misalignment in this study that 
uses disk-dominated galaxies. 

We conclude that angular momentum 
is roughly conserved by median factor $f_j^{'}\approx -0.21$ 
(corresponding to $j_{\rm s}/j_{\rm tot} \approx 0.62$) for 
disk-dominated central galaxies. A similar result is 
obtained in an independent analysis using the TNG100 run of \TNG\ 
\citep{Rodriguez-Gomez2022}. The overall 
correlation between galaxies and halos is maintained during the 
formation of disk-dominated galaxies. 
It is clear that the accretion of gas with high angular 
momentum dominates the growth of disk galaxies since $z=1.5$. 
Without experiencing violent mergers, the assembly of 
disky structures is able to conserve angular momentum 
as stars form from the cold gas. While biased collapse 
has been considered to play an important role in interpreting 
the observed $j_{\rm s}$-$M_{\rm s}$ relation 
\citep[][]{ShiJingjing2017, Posti2018a}, our results indicate 
that its effect has been largely erased in the local Universe. 

\begin{figure*}[htbp]
\begin{center} %
\includegraphics[width=\textwidth]{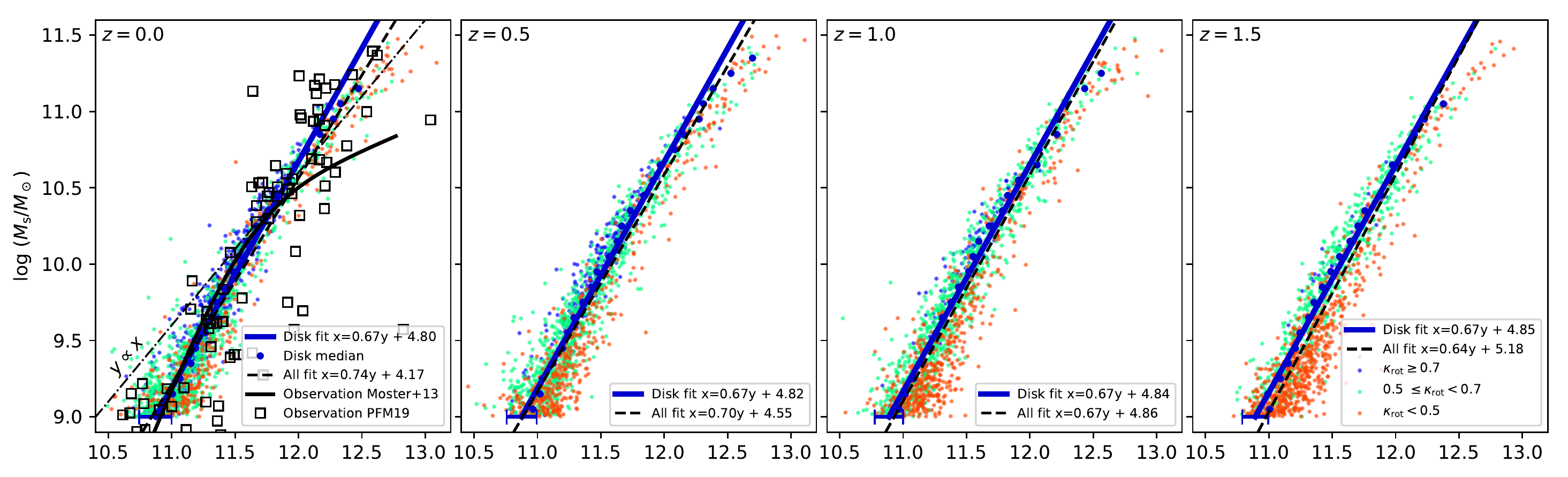}
\includegraphics[width=\textwidth]{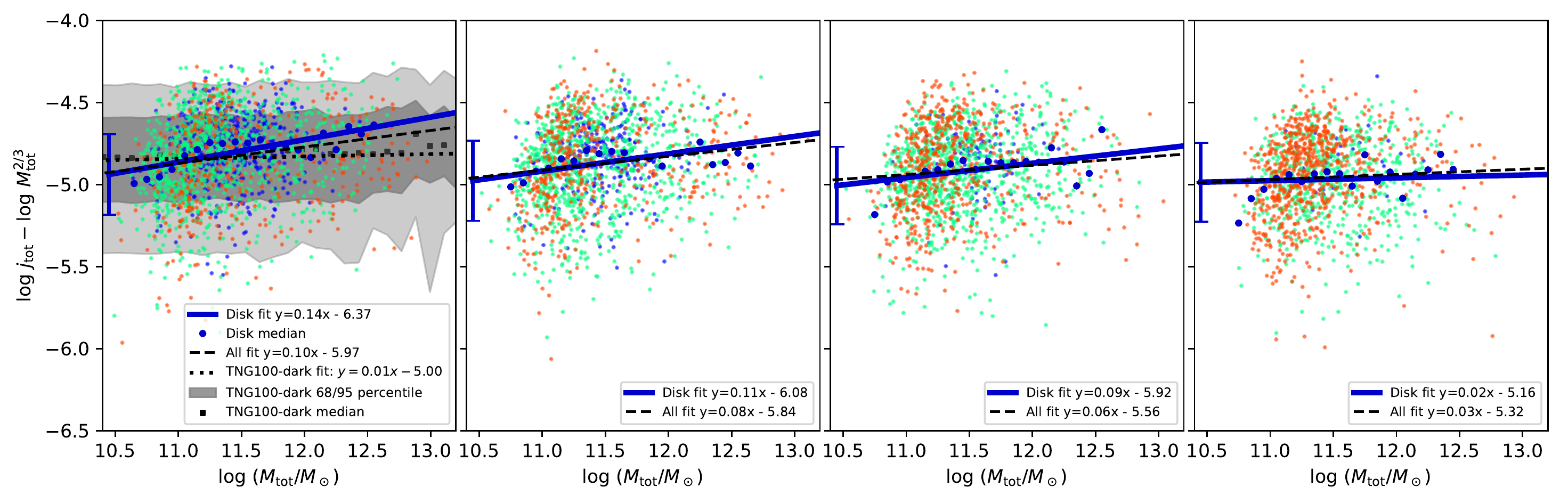}
\caption{Evolution of the $M_{\rm tot}$-$M_{\rm s}$ and $j_{\rm tot}$-$M_{\rm tot}$ relations of central galaxies from $z=1.5$ (right) to $z=0$ (left) in TNG50. The red, green, and blue symbols are central galaxies that correspond to spheroid-dominated galaxies, disk-dominated galaxies with $0.5 \leq \kappa_{\rm rot} < 0.7$, and disk-dominated galaxies with $\kappa_{\rm rot} \geq 0.7$, respectively. In the upper panels, we overlay the observations of disk galaxies from PFM19 and \citet{Moster2013} for comparison. Both the linear fitting and medians are measured using equal bins in log $M_{\rm s}$, thus giving the parameters $\beta$ and $f_{\rm m}^{'}$ of Equation \ref{eqMM} directly. In the bottom panels, we normalize $j_{\rm tot}$ by $M_{\rm tot}^{2/3}$ to highlight the discrepancy from tidal torque theory. In the bottom-left panel, the $j_{\rm tot}$-$M_{\rm tot}$ relation of central galaxies in TNG100-dark is overlaid for comparison. The shaded regions correspond to the 68 and 95 percentile envelopes. The dotted line and squares are the linear fitting result and median values, respectively.}
\label{fig:M0j0_onflow}
\end{center}
\end{figure*}

\subsection{Constraining the $j$-$M$ relation of halos with the SHMR}
\label{sec:SHMR}

According to the $j_{\rm s}$-$j_{\rm tot}$ relation, $\gamma \sim 1$, 
and therefore \refeq{eqmain} can be written as 
\begin{equation}\label{eqmain_v1}
\begin{aligned}
{\rm log}\ j_{\rm s} \simeq \alpha \beta\ {\rm log}\ M_{\rm s} + a + \alpha f_{\rm m}^{'} + f_j^{'},
\end{aligned}
\end{equation}
whose slope is determined by $\alpha$ and $\beta$. Tidal torque 
theory predicts $\alpha = 2/3$, which has been widely assumed. 
For this to hold, the index $\beta$ of the SHMR must be 
close to 1.

In the upper panels of \reffig{fig:M0j0_onflow}, we show the 
$M_{\rm s}$-$M_{\rm tot}$ relation of TNG50 
galaxies. The linear fitting (dash-dotted line) of 
disk-dominated galaxies (small blue and cyan dots) gives 
\begin{equation}\label{eq:M0Ms}
	\begin{aligned}
	{\rm log}\ M_{\rm tot} = (0.67\pm 0.01)\ {\rm log}\ M_{\rm s} - (4.80\pm 0.06)
	\end{aligned}
\end{equation} 
at $z=0$, according to which $\beta=0.67$, significantly smaller 
than 1. We overlay the SHMR measured by PFM19 (black squares), 
who estimated halo masses directly from the kinematics of 
extended HI in central 
disk galaxies. The SHMR of TNG50 galaxies matches well  
with the results of PFM19, while it is systematically 
offset from that derived from the abundance matching method 
\citep[e.g.,][the solid black profile]{Moster2013} for massive 
galaxies. The abundance matching method suggests that the
stellar-to-halo mass ratio follows a broken power-law 
relation peaking at $M_{\rm tot} \approx 10^{12} M_\odot$ 
\citep[][]{Wechsler&Tinker2018}, assuming there 
is no dependence on galaxy morphology. 
In relatively less massive galaxies with halo mass $< 10^{12} M_\odot$, 
both observations of PFM19 and simulations are consistent  
with abundance matching. The absence of a significant down-bending 
break in the SHMR of massive disk galaxies, however, challenges 
the results of abundance matching. While the SHMR of TNG50 
galaxies are roughly consistent with that of PFM19, we do 
see a relatively minor down-bending break for 
$M_{\rm tot} > 10^{12} M_\odot$ in TNG50. We have confirmed 
that the TNG100 simulation run in an 8 times larger box also exhibits 
a similar minor down-bending break.
A similar result was obtained in \citet{Marasco2020} using the 
TNG100 run. They suggested that the AGN feedback 
used in the TNG simulations is too efficient at suppressing 
star formation in massive disk galaxies. Here we simply 
use a linear fitting to describe 
the SHMR because (1) massive disk galaxies with 
$M_{\rm s} \geq 10^{11} M_\odot$ that are offset significantly 
are rare, and (2) the difference from the median values 
(large blue dots) is minor. 

The \TNG\ simulations suggest that there is a non-negligible 
correction to the $j \propto M^{2/3}$ relation when baryonic 
processes are considered. The lower panels of 
\reffig{fig:M0j0_onflow} show the $j_{\rm tot}$-$M_{\rm tot}$
relation. Using 
${\rm log}\ j_{\rm tot}- {\rm log}\ M_{\rm tot}^{2/3}$ as 
the $y$-axis to highlight 
the discrepancy from the traditional tidal torque theory, 
the bottom-left panel clearly shows that the dark 
matter-only runs in the \TNG\ simulations indeed generate 
$j \propto M^{2/3}$ (TNG100-dark, corresponding to the 
dotted line and shaded regions), consistent 
with the theoretical expectation. However, in the presence 
of baryons, the $j$-$M$ relation gradually deviates 
from this relation below $z=1$ (lower panels of
\reffig{fig:M0j0_onflow}). At $z=0$, fitting the central 
galaxies dominated by disks from TNG50 gives 
\begin{equation}
	\begin{aligned}
	{\rm log}\ j_{\rm tot} = (0.81\pm 0.02)\ {\rm log}\ M_{\rm tot} - (6.37\pm 0.21)
	\end{aligned}
\end{equation} 
The power-law index reaches $\alpha = 0.81$ at $z=0$. 
Combining with the SHMR of disk-dominated galaxies 
(equation \ref{eq:M0Ms}),  
\begin{equation}\label{eqmain_v2}
	\begin{aligned}
	{\rm log}\ j_{\rm s} \simeq 0.54\ {\rm log}\ M_{\rm s} + f_j^{'} - 2.48,
	\end{aligned}
\end{equation}
which explains perfectly the $j_{\rm s}$-$M_{\rm s}$ 
index of 0.55 of disk galaxies at $z=0$ (\reffig{fig:Msjs_evo}). 
Apparently, 
the decrease of $f_j^{'}$ leads to a parallel shift of the 
$j_{\rm s}$-$M_{\rm s}$ relation from disk-dominated toward more
spheroid-dominated galaxies following a nearly parallel 
sequence. Shown in \reffig{fig:fj}, $f_j^{'}\approx 0$ 
for the galaxies with $\kappa_{\rm rot} \geq 0.7$ at $z=0$. 
For all disk-dominated galaxies ($\kappa_{\rm rot} \geq 0.5$),  
for which $f_j^{'} \approx -0.21$, \refeq{eqmain_v2} gives 
${\rm log}\ j_{\rm s} \simeq 0.54\ {\rm log}\ M_{\rm s} - 2.69$, 
which predicts exactly the outcome of the 
$j_{\rm s}$-$M_{\rm s}$ relation of disk-dominated 
galaxies at $z=0$ (equation \ref{eqjsMs}). The mass ratio 
of the spheroidal component quantified by $\kappa_{\rm rot}$ 
clearly correlates inversely with $f_j^{'}$ (\reffig{fig:fj}), 
offering a qualitative explanation for the morphological 
dependence of the $j_{\rm s}$-$M_{\rm s}$ relation.

At high redshifts, the deviation from the local $j_{\rm s}$-$M_{\rm s}$ 
relation is partially explained by the evolution of the 
$j_{\rm tot}$-$M_{\rm tot}$ relation and the retention 
factor of angular momentum, as the $M_{\rm tot}$-$M_{\rm s}$ 
relation remains nearly invariant since $z=1.5$. The 
$j_{\rm tot}$-$M_{\rm tot}$ and 
$M_{\rm tot}$-$M_{\rm s}$ relations at $z=1.5$ give 
log $j_{\rm s} = 0.46\ {\rm log}\ M_{\rm s} - 1.86$ in the 
case of $j_{\rm tot}=j_{\rm s}$, which still cannot fully 
explain the index 0.34 of the $j_{\rm s}$-$M_{\rm s}$ relation 
at high redshifts. This may be due to the fact that galaxies 
have been affected by biased collapse and by losses of angular 
momentum due to gas-rich mergers and clumpy instabilities 
at $z>1.5$, as a consequence of which the retention factor 
$f_j^{'}$ is smaller at high redshifts (right-most panel of \reffig{fig:fj}).

Our results suggest that the dark matter-only $j\propto M^{2/3}$ 
relation cannot explain the $j_{\rm s}$-$M_{\rm s}$ relation. 
This is because the effect of central halos gaining angular 
momentum under the effect of baryonic processes needs to be 
included. The SHMR can be used to probe 
the $j$-$M$ relation in the local Universe when the effect of 
biased collapse becomes insignificant. It it worth 
emphasizing that all conclusions above are 
mainly drawn using less massive galaxies, whose SHMR can be 
described by a linear fit and is not sensitive to the 
down-bending break in massive ($M_{\rm tot} > 10^{12}M_\odot$) galaxies.

The mechanism responsible for the discrepancy from 
the traditional tidal torque theory is still not 
well known. In previous studies, \citet{Zjupa&Springel2017} 
suggested that the angular momentum in Illustris galaxies is
underestimated by dark matter-only simulations for 
especially massive cases. Shown in their Figure 19, the spin 
parameter has indeed a weak dependence on halo mass 
($M_{\rm h} > 10^{11}M_\odot$) in a similar manner to 
our halo $j$-$M$ relation. We have verified that 
the disk-dominated galaxies in the TNG100 run have 
a similar discrepancy to the cases in the TNG50 
\citep[see the result of TNG100 also in Figure 10 of][]{Rodriguez-Gomez2022}. 
\citet{ZhuQirong2017} 
showed that the presence of the baryonic component 
can induce net rotation in the inner regions 
of dark matter halos, which may lead to the 
increase of their angular momentum. \citet{Pedrosa2010} 
suggested that central galaxies may acquire 
angular momentum from their satellites that 
are disrupted by dynamical friction. Similarly, 
\citet{LuShengdong2022} showed that galaxy 
interactions can inject angular momentum to the 
circumgalactic medium. Moreover, galaxies with 
relatively lower $j_{\rm tot}$ may have higher 
probability of mergers, thus transforming 
their morphology into ellipticals. We see that the 
slope of all galaxies (black dashed lines) is slightly 
smaller, but it cannot fully explain the increase 
of $j_{\rm tot}$. 

\section{Summary}

In this paper, we show that the TNG50 simulation 
reproduces the observed scaling relation between stellar specific angular 
momentum $j_{\rm s}$ and mass $M_{\rm s}$ of galaxies, 
as measured in the local Universe. The disk-dominated 
central galaxies in TNG50 follow 
log $j_{\rm s} =0.55\ {\rm log}\ M_{\rm s} - 2.77$, 
which matches observations remarkably well.
Our result confirms that the observed 
$j_{\rm s}$-$M_{\rm s}$ relation may be regarded as 
evidence that the formation of disk galaxies is 
tightly correlated with dark matter halos. 
However, the theoretical $j$-$M$ relation ($j \propto M^{2/3}$) 
from dark matter-only simulations is not able to 
explain the $j_{\rm s}$-$M_{\rm s}$ relation. 

We show that the local $j_{\rm s}$-$M_{\rm s}$ 
relation develops at $z\lesssim 1$ in disk galaxies. 
During this epoch, disky structures form or grow 
significantly. Angular momentum is roughly conserved 
during the assembly of disky structures, which leads 
to a median retention factor ${\rm log}\ j_{\rm s}/j_{\rm tot} = -0.07_{-0.17}^{+0.15} \ (-0.21_{-0.24}^{+0.21})$ 
for disk-dominated galaxies with $\kappa_{\rm rot} \geq 0.7\ (0.5)$.
The $j_{\rm s}$-$M_{\rm s}$ relation of disk galaxies 
in the local Universe can be well explained by a simple 
model for which $j_{\rm tot}\propto M_{\rm tot}^{0.81}$, $M_{\rm tot}\propto M_{\rm s}^{0.67}$, 
and $j_{\rm s} \propto j_{\rm tot}$, where $j_{\rm tot}$ 
is the overall specific angular momentum and 
$M_{\rm tot}$ is the mass of the dark and baryonic components. 
Because of the cumulative accretion of mass with 
high angular momentum, the effect of biased collapse 
has been erased at low redshifts. The index 0.55 of the 
$j_{\rm s}$-$M_{\rm s}$ relation comes from the indices 
of the $j_{\rm tot}$-$M_{\rm tot}$ and 
$M_{\rm tot}$-$M_{\rm s}$ relation. We  
show that there is a non-negligible deviation from the 
halo $j\propto M^{2/3}$ relation to 
explain the $j_{\rm s}$-$M_{\rm s}$ relation.
This model further suggests that the stellar-to-halo 
mass ratio of disk 
galaxies increases monotonically following a nearly 
power-law function, which is consistent with the latest 
dynamical measurement of disk galaxies. This 
challenges the general expectation from 
abundance matching that the stellar-to-halo 
mass ratio of disk galaxies decreases toward the massive end. 
Moreover, the retention factor of angular momentum 
inversely correlates with the mass ratio of spheroids, 
which possibly leads to the morphological dependence 
of the $j_{\rm s}$-$M_{\rm s}$ relation. 

\begin{acknowledgements}
We thank an anonymous referee for helpful suggestions. The authors acknowledge constructive comments and suggestions from S. M. Fall, L. Posti, S. Liao, and J. Shi. LCH was supported by the National Science Foundation of China (11721303, 11991052, 12011540375) and the China Manned Space Project (CMS-CSST-2021-A04, CMS-CSST-2021-A06). MD and HRY acknowledge the support by the China Manned Space Program through its Space Application System, and the National Science Foundation of China 11903021 and 12173030. VPD was supported by STFC Consolidated grant ST/R000786/1. 
The TNG50 simulation used in this work, one of the flagship runs of the IllustrisTNG project, has been run on the HazelHen Cray XC40-system at the High Performance Computing Center Stuttgart as part of project GCS-ILLU of the Gauss centres for Supercomputing (GCS). The authors are acknowledged for the help with the high-performance computing of Xiamen University. This work is also supported by the High-performance Computing Platform of Peking University, China. 

\end{acknowledgements}

\end{document}